\documentclass[a4paper]{article}

\usepackage{amsmath}
\usepackage{mathtools}
\usepackage{amsthm}
\usepackage{amssymb}
\usepackage{graphicx}
\usepackage{epstopdf}
\DeclareMathOperator{\sech}{sech}

\def\ii{\int\limits_{\mathbb{R}}}
\def\jj{\int\limits_{B(x)}^{\eta(x)}}

\begin{document}

\begin{center}

\section*{\LARGE \bf{Hamiltonian models for the propagation of irrotational surface gravity waves over a variable bottom}}

\vskip1cm

{\large \bf Alan Compelli$^{a,b,\dag}$},  {\large \bf  Rossen I. Ivanov$^{a,b,\ddag}$} and \\ {\large \bf Michail D. Todorov$^{c,\clubsuit}$} 

\vskip1cm

\hskip-.3cm
\begin{tabular}{c}
\\
$^{a}${\it School of Mathematical Sciences, Dublin Institute of Technology, }\\ {\it Kevin Street, Dublin 8, Ireland} \\
\\
$^{b}${\it Erwin Schr\"odinger Int. Institute for Mathematics and Physics, }\\ {\it University of Vienna, Vienna, Austria,} \\
\\
$^{c}${\it Department  of Differential Equations,}\\{\it
Faculty of Applied Mathematics and Informatics, }\\ {\it Technical University of Sofia, 8 Kl. Ohridski Blvd., 1000 Sofia, Bulgaria} 
\\
\\
\\{\it $^\dag$e-mail: alan.compelli@dit.ie}
\\{\it $^\ddag$e-mail: rossen.ivanov@dit.ie  }
\\{\it $^\clubsuit$e-mail: mtod@tu-sofia.bg    }
\\
\end{tabular}
\end{center}

\vskip0.5cm

\input epsf







\begin{abstract}
A single incompressible, inviscid, irrotational fluid medium bounded by a free surface and varying bottom is considered. The Hamiltonian of the system is expressed in terms of the so-called Dirichlet-Neumann operators. The equations for the surface waves are presented in Hamiltonian form. Specific scaling of the variables is selected which leads to approximations of Boussinesq and KdV types taking into account the effect of the slowly varying bottom.
The arising KdV equation with variable coefficients is studied numerically when the initial condition is in the form of the one soliton solution for the initial depth.
\end{abstract}

{\bf Mathematics Subject Classification (2010):} 35Q53, 35Q35, 37K05, 37K40
\\
\\
{\bf Keywords:} Dirichlet-Neumann Operators, KdV equation, Water waves, Solitons



\section{Introduction}
In 1968 V. E. Zakharov in his work \cite{Zakharov} demonstrated that the equations for the surface waves of a deep inviscid irrotational water have a canonical Hamiltonian formulation. The result has been extended to models with finite depth and flat bottom \cite{CraigGroves,CraigSulem}, for internal waves between layers of different density \cite{CraigGuyenneKalisch} as well as waves with added shear for constant vorticity \cite{CIP, CI, CIM, CompelliIvanov1,CompelliIvanov2}. A multilayer model based on the Green-Naghdi approximation has been proposed in \cite{CHP}.

The Hamiltonian approach to the wave motion has been extended by inclusion of variations of the bottom surface in several papers \cite{CraigGuyenneNichollsSulem,CraigGuyenneSulem,BCDGS}.
The Hamiltonian framework allows for approximations taking into account different scales when considering shallow water and long-wave regimes. As a result the arising model equations are of the type of well known shallow water equations like the Korteweg-de Vries (KdV) and Boussinesq equations.

The problem of waves with a variable bottom has a long history. In the pioneering work of Johnson \cite{Johnson71} a perturbed KdV equation is derived as a model for surface waves from Euler's governing equations for irrotational inviscid fluid (cf. also with \cite{Kaku}). The problem has been studied further by Johnson \cite{Johnson71,Johnson73} and several other authors, e.g. in \cite{Kaup78,KN80,KN85,KN85a}.

Boussinesq-type models with variable bottom have been derived and analysed in \cite{Nach1, Nach2}.

In this review paper we consider the scaling regime adopted in \cite{Johnson71} within the Hamiltonian framework of waves under the variable bottom of Craig \emph{et al} \cite{CraigGuyenneNichollsSulem,CraigGuyenneSulem,BCDGS}.  We use numerical solutions of the Johnson equation in order to analyse the effects of the propagation of solitary waves over a variable bottom.

\section{Setup}
\begin{figure}[h!]
\begin{center}
\fbox{\includegraphics[totalheight=0.23\textheight]{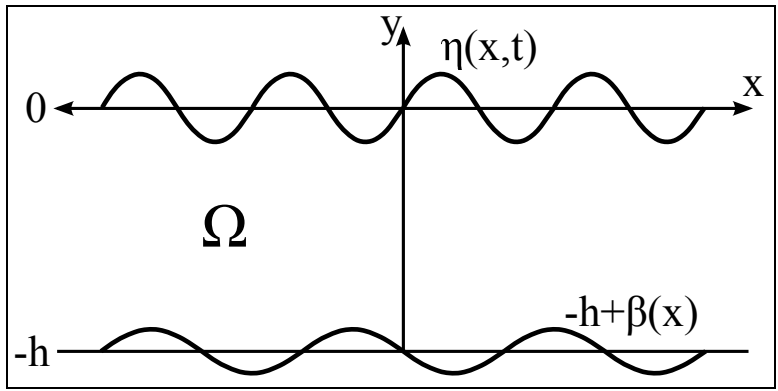}}
\caption{The system under study.}
\label{fig:thesisfigure_systemG}
\end{center}
\end{figure}

An inviscid, incompressible system is presented with a surface wave and variable bottom as shown in Fig. \ref{fig:thesisfigure_systemG}.

The average water surface is at $y=0$ and the wave elevation is given by the function $\eta(x,t).$  Therefore we have
\begin{equation}
\int_{\mathbb{R}} \eta(x,t) \, dx =0.
\end{equation}
The average bottom depth is at $y=-h$ and the bottom elevation from the average depth is given by the function $\beta(x).$ Hence the function $B(x)$ that represents the local depth $y=B(x)$ is
\begin{equation}
B(x)=-h+\beta(x).
\end{equation}
The body of the fluid which occupies the domain $\Omega$ is defined as
\begin{equation}
\label{Omega_1_VarBott}
\Omega:=\{(x,y)\in \mathbb{R}^2: B(x)<y<\eta(x,t)\}.
\end{equation}

The subscript notation $s$ will be used to refer to valuation on the surface $y=\eta(x,t)$ and $b$ will refer to valuation on the variable bottom $y=B(x)$.

\section{Governing equations}
Let us introduce the velocity field ${\bf{v}}=(u,v,0).$ The incompressibility $u_x+v_y=0$ and the irrotationality of the flow $u_y-v_x=0$ allow the introduction of a stream function $\psi$ and velocity potential $\varphi$ as follows:
\begin{equation}
\label{definitionofpsi_SystemG}
\begin{dcases}
    u=\psi_y=\varphi_x\\
    v=-\psi_x=\varphi_y.
\end{dcases}
\end{equation}

\noindent In addition,  $$\Delta \varphi=0, \qquad \Delta \psi =0 $$ in $\Omega.$ This leads to $$ |\nabla \varphi|^2 = \nabla \cdot (\varphi \nabla \varphi)= \text{div} (\varphi \nabla \varphi).  $$

The Euler equations written in terms of the velocity potential produce the Bernoulli condition on the surface
\begin{equation}
\label{Bernoulli}
(\varphi_t)_s+\frac{1}{2}|\nabla \varphi|_s^2+g \eta=0
\end{equation}
where $g$ is the acceleration due to gravity.

There is a kinematic boundary condition on the wave surface
\begin{equation}
\label{KBC}
v=\eta_t +u\eta_x \quad \text{or}  \quad (\varphi_y)_s=\eta_t + (\varphi_x)_s \eta_x
\end{equation}
and on the bottom
\begin{equation}
\label{KBC_VarBott}
(\varphi_y)_b=B_x(\varphi_x)_b.
\end{equation}

We make the assumption that all considered functions $\eta(x,t)$, $\varphi(x,y,t)$, $\beta(x)$ are in the Schwartz class for the $x$ variable, that is declining fast enough when $x\to \pm \infty$ (for all values of the other variables). In other words we describe the propagation of solitary waves.

\section{Hamiltonian formulation}
The Hamiltonian of the system will be represented as the total energy of the fluid

\begin{equation}
H=\frac{1}{2}\rho \int\int_{\Omega} (u^2+v^2)dy dx+\rho g \int \int _{\Omega} y dydx.
\end{equation}
It can be written in terms of the variables $(\eta(x,t),\varphi(x,y,t))$ as
\begin{equation}
H[\eta,\varphi]= \frac{1}{2}\rho\ii  \jj |\nabla \varphi|^2 dydx
+\rho g\ii\jj y dydx.
\end{equation}

We introduce as usual the variable $\xi$ proportional to the potential evaluated on the surface
\begin{equation}
\xi(x,t):=\rho \varphi(x,\eta(x,t),t)\equiv \rho \phi(x,t),
\end{equation}

\noindent and the Dirichlet-Neumann operator $G(\beta,\eta)$ given by
\begin{equation}
 G(\beta,\eta)\phi =-\eta_x (\varphi_x)_s+(\varphi_y)_s =(-\eta_x, 1) \cdot (\nabla \varphi)_s=\sqrt{1+\eta_x^2}\,\,  {\bf n}_s \cdot (\nabla \varphi)_s
\end{equation}

\noindent where  ${\bf n}_s = (-\eta_x, 1)/\sqrt{1+\eta_x^2}$ is the outward-pointing unit normal vector (with respect to $\Omega$) to the wave surface.

With Green's Theorem (Divergence Theorem) the Hamiltonian can be written as
\begin{equation}
\label{mainHamiltonian}
H[\eta, \xi]=\frac{1}{2\rho}\int_{\mathbb{R}} \xi G(\beta,\eta) \xi dx
+\frac{1}{2}\rho g\int_{\mathbb{R}} (\eta^2-B^2) dx .
\end{equation}

On the bottom the outward-pointing unit normal vector is 

\noindent ${\bf n}_b=(B_x, -1)/\sqrt{1+B_x^2}$ and   $$\sqrt{1+B_x^2}\, \, {\bf n}_b \cdot (\nabla \varphi)_b =(B_x, -1) \cdot \left(( \varphi_x)_b, (\varphi_y)_b\right)=(B_x( \varphi_x)_b- (\varphi_y)_b)=0  $$ thus $  {\bf n}_b \cdot (\nabla \varphi)_b =0$ and no bottom-related terms are present. Noting that the term $\int_{\mathbb{R}} B^2(x) dx$ is a constant and will not contribute to $\delta H$, we renormalise the Hamiltonian to

\begin{equation}
\label{mainHam}
H[\eta, \xi]=\frac{1}{2\rho}\int_{\mathbb{R}} \xi G(\beta,\eta) \xi dx
+\frac{1}{2}\rho g\int_{\mathbb{R}} \eta^2 dx.
\end{equation}

The variation of the Hamiltonian can be evaluated as follows. Green's Theorem transforms the following expression to contributions from the surface and the bottom:

\begin{align}
\delta\bigg[\rho\int \int _{\Omega_{}} |\nabla \varphi|^2 dydx\bigg]&=\rho\int_{\mathbb{R}} \big((\varphi_y)_s-(\varphi_x)_s \eta_x\big)(\delta\varphi)_s dx\notag\\
&\quad-\rho\int_{\mathbb{R}} \big((\varphi_y)_b-(\varphi_x)_b B_x\big)(\delta\varphi)_b dx
+\frac{1}{2}\rho\int_{\mathbb{R}}  |\nabla \varphi|_s^2\, \delta \eta \,dx.
\end{align}

\noindent Clearly,
\begin{align}
\delta\bigg[\rho g\int_{\mathbb{R}} \eta^2 dx\bigg]&=2\rho g\int_{\mathbb{R}} \eta\delta\eta dx.
\end{align}

\noindent Noting that the variation of the potential on the wave surface is given as
\begin{equation}
(\delta\varphi)_s=\delta\phi-(\varphi_y)_s\delta\eta, \qquad (\delta\varphi)_b=\delta\phi_b
\end{equation}
where
\begin{equation}
\phi(x,t) :=\varphi(x,\eta(x,t),t), \qquad \phi_b(x,t) :=\varphi(x,B(x),t)
\end{equation}
we write
\begin{multline}
\delta H=\rho\int_{\mathbb{R}} \big((\varphi_y)_s-(\varphi_x)_s \eta_x\big)\big(\delta\phi-(\varphi_y)_s\delta\eta\big) dx
-\rho\int_{\mathbb{R}} \big((\varphi_y)_b-(\varphi_x)_b B_x\big)\delta\phi_b dx\\
+\frac{1}{2}\rho\int_{\mathbb{R}}  |\nabla \varphi|_s^2\, \delta \eta \,dx
+\rho g\int_{\mathbb{R}} \eta\delta\eta dx.
\end{multline}

\noindent It is noted from (\ref{KBC_VarBott}) that
\begin{equation}
\frac{\delta H}{\delta \phi_b}=-\rho((\varphi_y)_b- B_x(\varphi_x)_b) =0,
\end{equation}
which represents the bottom condition in variational form. It also explains the fact that $H$ does not depend on  $\phi_b$. Evaluating $\delta H/ \delta\xi$ we remember that $\rho\delta\phi=\delta\xi$ and therefore

\begin{equation}
\frac{\delta H}{\delta \xi}=(\varphi_y)_s-(\varphi_x)_s \eta_x=\eta_t
\end{equation}
\noindent due to (\ref{KBC}).

\noindent Next we compute
\begin{equation}
\frac{\delta H}{\delta \eta}= - \rho \big((\varphi_y)_s-(\varphi_x)_s \eta_x\big)(\varphi_y)_s    +\frac{1}{2} \rho |\nabla \varphi|_s^2  +\rho g \eta.
\end{equation}

\noindent Noting that, using the kinematic boundary condition (\ref{KBC}),
\begin{align}
 -  \big((\varphi_y)_s-(\varphi_x)_s \eta_x\big)(\varphi_y)_s
 & = - \eta_t  (\varphi_y)_s
\end{align}
we write
\begin{equation}
\frac{\delta H}{\delta \eta}= - \rho \eta_t(\varphi_y)_s    +\frac{1}{2}\rho  |\nabla \varphi|_s^2  +\rho g \eta.
\end{equation}

\noindent Recall that $$ \xi_t = \rho((\varphi_t)_s   +(\varphi_y)_s \eta_t), $$

\noindent then
%
\begin{align}
\frac{\delta H}{\delta \eta}&= - \xi_t + \rho  \left( (\varphi_t)_s   +\frac{1}{2}  |\nabla \varphi|_s^2+ g \eta \right)= - \xi_t
\end{align}

\noindent by the virtue of the Bernoulli equation (\ref{Bernoulli}).

Thus we have canonical equations of motion

\begin{equation}
\label{EOM}
    \xi_t= -\frac{\delta H}{\delta \eta}, \qquad
    \eta_t= \frac{\delta H}{\delta \xi}
\end{equation}

\noindent where the Hamiltonian is given by \eqref{mainHamiltonian}. Introducing the variable $\mathfrak{u}=\xi_x$ we can represent \eqref{EOM} in the form

\begin{equation}
\label{EOM1}
    \mathfrak{u}_t= -\left(\frac{\delta H}{\delta \eta}\right)_x, \qquad
    \eta_t= - \left(\frac{\delta H}{\delta \mathfrak{u}}\right)_x.
\end{equation}

\section{Taylor expansion of the Dirichlet-Neumann operator}
We introduce some basic properties of the Dirichlet-Neumann operator. The details can be found in \cite{CraigGuyenneNichollsSulem, BCDGS}. The operator can be expanded as
\begin{equation}
G(\beta,\eta)=\sum_{j=0}^{\infty} G^{(j)}(\beta,\eta)
\end{equation}
where $ G^{(j)}(\beta,\eta)\sim (\eta/h)^j.$  The surface waves are assumed small, i.e. $|\eta_{\mathrm{max}}|/h\ll 1$ and one can expand with respect to $|\eta/h |\ll 1 $ as follows: \begin{equation}
G^{(0)}=D\tanh(hD) + DL(\beta), \qquad G^{(1)}=D\eta D -G^{(0)}\eta G^{(0)}.
\end{equation}

The operator $D=-i\partial/\partial x$ has an eigenvalue $k=2\pi/\lambda$ for a given wavelength $\lambda$, when acting on monochromatic plane wave solutions proportional to $e^{ik(x-c(k)t)}.$ In the long-wave case $hD$ has an eigenvalue $$hk=\frac{2\pi h}{\lambda} \ll 1 $$ thus when $h k$ is small, one can formally expand in powers of $h D$. 

For the operator $L(\beta)$ the expansion is
\begin{equation}
L(\beta)=\sum_{j=0}^{\infty} L_j(\beta)
\end{equation}

\noindent where $L_j \sim (\beta/h) ^j$. It is assumed that $\beta(x)$ can be of the order of $h$ provided that the bed stays away from the surface, that is $|\beta_{\text{max}}|/h$ of $\mathcal{O}(1)$ with $|\beta(x)|/h < 1.$ This way one can expand in powers of $\beta/h .$  Using the recursive formulae in \cite{CraigGuyenneNichollsSulem},
\begin{align}
L_0&=0\notag\\
L_1&=-\sech{(hD)}\beta D\sech{(hD)}\notag\\
L_2&=-\sech{(hD)}\beta D \sinh{(hD)}\sech{(hD)}\beta D \sech{(hD)}\notag\\
L_3&=-\sech{(hD)}\Bigg(\frac{\beta^3}{3!}\sech{(hD)}D^3+ \frac{\beta^2}{2!}D^2\cosh{(hD)}L_1
- \frac{\beta}{1!}D\sinh{(hD)}L_2 \Bigg).\notag
\end{align}

The Dirichlet-Neumann operator is   \begin{multline}
G(\beta,\eta)=D\tanh(hD)+DL(\beta)+D \eta  D
- D\tanh(hD)  \eta D\tanh(hD) \\
- D\tanh(hD)  \eta  DL(\beta)
- DL(\beta) \eta  D\tanh(hD)
- DL(\beta) \eta  DL(\beta) +\mathcal{O}((\eta/h)^2).
\end{multline}

\noindent Keeping expansions up to $(hD)^4,$ $(\eta/h)^1(hD)^2$ and $(\beta/h)^3$ in $G(\beta, \eta)$ we need
\begin{align}
\tanh(hD)&=hD-\frac{1}{3}h^3D^3+\mathcal{O}((hD)^5)\notag\\
L_1&=-\beta D+\frac{1}{2}h^2\beta D^3+\frac{1}{2} h^2D^2\beta D+\mathcal{O}((hD)^4)\notag\\
L_2&=-h\beta D^2\beta D+\mathcal{O}((hD)^4)\notag\\
L_3&=-\frac{1}{6}\beta^3D^3+ \frac{1}{2}\beta^2D^2\beta D+ \mathcal{O}((hD)^4)\notag
\end{align}
and hence, the truncated expansion is
\begin{multline}
G(\beta, \eta) =\\
D\left(h-\beta -\frac{h^3}{3}D^2 +\frac{h^2}{2} \beta D^2+\frac{h^2}{2}  D^2\beta  -h\beta D^2\beta -\frac{1}{6}\beta^3 D^2+ \frac{1}{2}\beta^2D^2\beta  + \eta \right)D \\
 +\mathcal{O}(\left(\frac{\eta}{h}\right)^2,\left(\frac{\eta}{h}\right)(h D)^2,(hD)^4).
\end{multline}

\noindent In the case when $\beta= \beta(\varepsilon x)$ with $\varepsilon$ of order $(hD)^2$ (or smaller) the commutator of $\beta$ and $D$ is proportional to $\varepsilon \beta'(\varepsilon x)$ which is of order $(hD)^2$ (or smaller) and therefore small with respect to $\beta D$ or $D \beta$. Therefore we take the truncated expansion

\begin{equation}
G(b, \eta)=D\left((h-\beta) -\frac{1}{3}D(h-\beta)^3 D  + \eta \right)D =D\left( b(X) -\frac{1}{3}D b^3(X) D  + \eta \right)D  , \label{DN_1}
\end{equation}

\noindent where $b(X)= h-\beta(\varepsilon x)$ is the local depth and $X=\varepsilon x$ indicates that the bottom depth varies slowly with $x$.

\section{Boussinesq and KdV approximations}

We introduce, as usual, the small scale parameters
\begin{equation}
\varepsilon=\frac{a}{h}\quad\mbox{and}\quad\delta=\frac{h}{\lambda},\notag
\end{equation}
where $a$ is the wave amplitude and $\lambda$ is the wavelength, and consider the long-wave and shallow water scaling regime. We consider the wave propagation regime with $\eta/h$ of order $\varepsilon$; $h D$ with its eigenvalue  $ hk$ of order $\delta. $ The quantity $\mathfrak{u}=\xi_x$ has the magnitude of a velocity (multiplied by $\rho$) and therefore is of order $\varepsilon$; $b(X)$ is of order 1 and most importantly, $\varepsilon \sim \delta^2$ which usually leads to the Boussinesq and KdV propagation regimes.

The magnitudes of the quantities can be made explicit by the change

\begin{equation} \frac{\eta}{h} \rightarrow \varepsilon \frac{\eta}{h}, \qquad hD \rightarrow \delta hD, \qquad \mathfrak{u}\rightarrow  \varepsilon \mathfrak{u}, \end{equation} where now all quantities like $\eta,$ $\mathfrak{u}$ and $b(X)$ are $\mathcal{O}(1)$. We write the Hamiltonian (\ref{mainHam}) in terms of the scaled variables, using the expansion for the Dirichlet-Neumann operator given in (\ref{DN_1}), and keeping only terms of order $\varepsilon ^3$:
\begin{equation}
H[\mathfrak{u}, \eta]=\frac{\varepsilon^2}{2\rho}\int_{\mathbb{R}}  \mathfrak{u}\left(b(X) -\delta^2\frac{1}{3}D b^3(X) D  + \varepsilon \eta \right) \mathfrak{u} dx
+\frac{1}{2}\varepsilon^2\rho g\int_{\mathbb{R}} \eta^2 dx + \mathcal{O}(\varepsilon^4).
\end{equation}

\noindent As $\varepsilon^2$ is an overall scale factor we can work with the rescaled Hamiltonian
\begin{equation}
H[\mathfrak{u}, \eta]=\frac{1}{2\rho}\int_{\mathbb{R}}  \mathfrak{u}\left(b(X) -\delta^2\frac{1}{3}D b^3(X) D  + \varepsilon \eta \right) \mathfrak{u} dx
+\frac{1}{2}\rho g\int_{\mathbb{R}} \eta^2 dx. \label{HH}
\end{equation}

The Boussinesq-type equations of motion, truncating at $\mathcal{O}(\varepsilon)=\mathcal{O}(\delta^2) $, obtained from \eqref{HH} and \eqref{EOM1} and the commutation of $b$ and $\partial_x$ in the terms of orders $\varepsilon$ and $ \delta^2$ are
\begin{equation}
\label{LWEOM}
\begin{dcases}
    \rho \eta_t=-(b\mathfrak{u})_x-\frac{\delta^2}{3}  b^3 \mathfrak{u} _{xxx}-\varepsilon (\eta  \mathfrak{u})_x \\
    \mathfrak{u}_t=-\rho g \eta_{x}-\varepsilon\frac{1}{\rho}\mathfrak{u}\mathfrak{u}_{x}.
\end{dcases}
\end{equation}

In the case of constant depth $b=h$=const ( $g$, $\rho$, $\varepsilon$ and $\delta$ can be scaled out) the system becomes  \begin{equation}  \begin{dcases}
  & \eta_t+h\mathfrak{u}_x+\frac{1}{3}  h^3 \mathfrak{u} _{xxx}+(\eta  \mathfrak{u})_x=0  \nonumber \\
    & \mathfrak{u}_t +\eta_{x}+\mathfrak{u}\mathfrak{u}_{x}=0. \nonumber
    \end{dcases}
\end{equation}

\noindent This system is integrable, known as  the Kaup - Boussinesq system. Indeed, the transformation $\eta \to \eta - h$ eliminates the $hu_x$ term. A further transformation $(\partial_t, \partial_x)\to \gamma (\partial_t, \partial_x)$ with $\gamma^2=\frac{3}{4h^3}$ brings the system to the form  \begin{equation}
 \begin{dcases}
  &\eta_t+\frac{1}{4} \mathfrak{u} _{xxx}+(\eta  \mathfrak{u})_x=0  \nonumber \\
    &\mathfrak{u}_t+\eta_{x}+\mathfrak{u}\mathfrak{u}_{x}=0, \nonumber
    \end{dcases}
\end{equation}

\noindent which has a scalar Lax pair representation (due to D.J. Kaup, \cite{K75}) with a spectral parameter $\zeta$:

\begin{equation}
\begin{dcases}
  &\Psi_{xx}=\left( \left( \zeta - \frac{\mathfrak{u}}{2}\right)^2 -\eta \right) \Psi  \nonumber \\
   & \Psi_t=-\left(\zeta +\frac{\mathfrak{u}}{2}\right) \Psi_x +\frac{1}{4}\mathfrak{u}_x \Psi . \nonumber
\end{dcases}
\end{equation}

\noindent The solitons of the KB system have been studied by several authors, see for example \cite{K75,IL, HL} and the references therein. 

 Returning back to \eqref{LWEOM}, it is noted that the leading order terms satisfy the system of equations

\begin{equation}
\label{LOE}
\begin{dcases}
    \rho \eta_t=-b \mathfrak{u}_x \\
    \mathfrak{u}_t=-\rho g \eta_{x},
\end{dcases}
\end{equation}

\noindent or $$ \eta_{tt} - gb(X) \eta_{xx} =0,$$

\noindent which shows that $\eta$ satisfies the wave equation for a wavespeed $c\approx\sqrt{gb(X)}$, that is the wavespeed is nearly constant, depending on the slowly varying variable $X$. Thus, in the leading order, the initial disturbance $\eta(x,0)\equiv \eta_0(x)$ propagates (to the right) with a slowly varying speed $c(X)\approx\sqrt{gb(X)}$:  \begin{equation}
\label{zerosol}
 \eta(x,t)=\eta_0(x-c(X)t).
\end{equation}

\noindent In the leading order approximation also

\begin{equation}
\label{u_zero_ord}
 \mathfrak{u}(x,t)=\frac{\rho c(X)}{b(X)}\eta(x,t).
\end{equation}

The observations from the leading order approximation indicate that the so-called {\it slow} variables, like the characteristic $x-c(X)t,$ might be more adequate for the analysis of the model. Following Johnson \cite{Johnson71, Johnson} we select a slow variable of the form

\begin{equation}
\theta=\frac{1}{\varepsilon}R(X)-t,
\end{equation}
reminiscent of the {\it far field}  variable $\theta= x- t$. The function $R(X)$ will be determined in what follows so that $\theta$ will be the characteristic for the right running wave. As a next step, the equations will be written in terms of the slow variables $X, \theta$. The derivatives are related as follows:

\begin{equation}
\begin{dcases}
    \partial_x=R'(X)\partial_{\theta}+\varepsilon\partial_X\\
    \partial_t=-\partial_{\theta}.
\end{dcases}
\end{equation}

\noindent The operator $D$ is hence
\begin{equation}
\label{D_ito_theta}
D=-i\partial_x=-i(R'(X)\partial_{\theta}+\varepsilon\partial_X).
\end{equation}

\noindent Note that the leading order approximation in the new variables from (\ref{LWEOM}) is

\begin{equation}
\begin{dcases}
   \rho \eta_{\theta}= R'(X)b(X)  \mathfrak{u}_{\theta}\\
   \mathfrak{u}_{\theta}=\rho g R'(X)  \eta_{\theta}.
\end{dcases}
\end{equation}

\noindent giving $(R'(X))^2 g b(X) =1,$ or $R'(X)=\pm \frac{1}{\sqrt{g b(X)}},$  so we define
 \begin{equation}
\label{Rinv_def}
R'(X)=\frac{1}{\sqrt{g b(X)}}=\frac{1}{c(X)}.
\end{equation}
Hence we write the equations \eqref{LWEOM} in the new variables
\begin{equation}
\label{SW_Eq}
\begin{dcases}
   \rho \eta_{\theta}=\frac{b}{c}\mathfrak{u}_{\theta}+ \varepsilon b \mathfrak{u}_X
   +\delta^2\frac{b^3}{3c^3}\mathfrak{u}_{\theta \theta \theta}+\varepsilon\frac{1}{c}(\eta \mathfrak{u})_{\theta}\\
  \mathfrak{u}_{\theta}={\frac{\rho g}{c }} \eta_{\theta}+\varepsilon \rho g \eta_X +\varepsilon \frac{1}{\rho c}\mathfrak{u}\mathfrak{u}_{\theta}.
\end{dcases}
\end{equation}

\noindent We substitute $\mathfrak{u}_{\theta}$ from the second equation in (\ref{SW_Eq}) into the first one. In the terms of orders $\varepsilon$ and $\delta^2$ we substitute  $\mathfrak{u}$  with its leading order approximation \eqref{u_zero_ord}. The equation for $\eta$ is
\begin{equation}
\label{eta1}
 \varepsilon (2 c \eta_X + c_X \eta)+  \varepsilon\frac{3g}{c^2} \eta \eta_{\theta}+\delta^2\frac{c^2}{3g^2}\eta_{\theta \theta \theta}=0.
\end{equation}

\noindent We observe that all terms are of smaller orders $\varepsilon$ and $\delta^2$. Rescaling appropriately the variables to remove the scale parameters $\varepsilon$ and $\delta^2$ and introducing the variable $E(\theta,X)=\sqrt{c(X)}\eta(\theta,X)$ we write the equation in (\ref{LWEOM}) in the form of a KdV equation with variable coefficients:

\begin{equation} \label{KdVa}
2E_X+\frac{3g}{c^{7/2}}E E_{\theta }+ \frac{c}{3g^2} E_{\theta\theta\theta}=0.
\end{equation}

\noindent This is exactly the equation obtained by Johnson  \cite{Johnson71} via appropriate expansions from the governing equations of the fluid motion, (see also the derivation in \cite{Johnson}) so we can refer to it as {\it Johnson's equation}.

In the propagation regime where $\delta^2 \ll \varepsilon$ clearly the $\delta^2$-term in \eqref{eta1} has to be neglected and one obtains the dispersionless Burgers equation

\begin{equation} \label{B}
2E_X+\frac{3g}{c^{7/2}}E E_{\theta }=0.
\end{equation}

\noindent This equation does not have globally smooth solutions, its solutions always form a vertical slope and break.

\section{The Johnson equation and soliton propagation}

The Johnson model \eqref{eta1} has been studied previously in \cite{Johnson73,KN80,KN85,KN85a,Kaup78}. It has been noted that a change of variables $$X\rightarrow \tilde{X}=\frac{1}{6g^{3/2}}\int \sqrt{b(X)} \, d X $$ and rescaling of $\eta \rightarrow -\frac{2c^4}{3g^3} \eta$ transforms the equation to the form of a perturbed KdV equation

\begin{equation}
\label{eta2}
 \varepsilon  \eta_{\tilde{X}} -  \varepsilon 6 \eta \eta_{\theta}+\delta^2\eta_{\theta \theta \theta}=-\varepsilon \Gamma \eta
\end{equation}  where the perturbation is on the right hand side with $$ \Gamma(X)= \frac{3\sqrt{g}}{2b^{3/2}} b'(X). $$

\noindent The perturbed KdV equation can be treated within the framework of the inverse scattering approach for the KdV equation. The basics of this approach are outlined for example in \cite{Lamb, Kaup78, GI}. If the initial condition is a pure soliton solution for the KdV equation, due to the perturbation, waves of radiation will appear and will decrease the energy of the initial soliton. The soliton perturbation theory however is unwieldy and in the case when the perturbation gives birth to new solitons is even more problematic. In our study, as we shall see below, new solitons are born when the initial soliton travels over an uneven bottom.

The model equation \eqref{KdVa} can be written in terms of the local depth $b(X)$, that is $E=b^{1/4} \eta$ (taking for simplicity $g=1$):

\begin{equation} E_X + \frac{3}{2}[b(X)]^{-7/4}E  E_{\theta} +\frac{1}{6}[b(X)]^{1/2}E_{\theta \theta \theta}=0. \label{KdVt1}\end{equation}

\noindent We observe that $X$ plays the role of the time-like variable in the usual KdV setting and $\theta$ - the space-like variable. In order to keep this analogy in what follows we replace $X$ with $\tau,$ i.e. $X\equiv \tau$ while $\theta$ is a space-like variable. This surprising outcome is due to the fact that the original $(x,t)$ variables are both of order $1/\varepsilon$ (while combinations like the characteristics $x-c(X) t$ are of order 1) and at leading order both $x$ and $t$ can measure time, see also the explanation in \cite{Johnson}. Thus we consider now the KdV with ``time''-dependent coefficients
\begin{equation} E_{\tau} + \frac{3}{2}[b(\tau)]^{-7/4}E  E_{\theta} +\frac{1}{6}[b(\tau)]^{1/2}E_{\theta \theta \theta}=0. \label{tdkdv}\end{equation}

\noindent According to the classification of the KdV equations with variable coefficients, \eqref{tdkdv} does not appear to be transformable to the KdV equation for any choice of $b(\tau)$, see for example \cite{OV}.  We proceed by specifying the $\tau$-dependent function \begin{equation} b(\tau)=h_0(1-\alpha \tanh (\beta \tau)) \label{b} \end{equation} where $h_0,$ $\beta$ and $0<\alpha <1$ are appropriate constants. This function represents a ramp at $\tau=0$ between two constant values (two depths), $\beta$ describes the steepness of the ramp at $\tau=0$. Thus, the initial condition will be taken at $ \tau_0 <0 $ before the ramp is switched on at $\tau=0$.  The initial profile then is taken as the KdV 1-soliton when $b$ is constant: ($\tanh \beta \tau \approx 1$) and  $b_0\approx h_0(1+\alpha);$  
\begin{equation}  E_-(\theta,\tau_0)=A \text{sech} ^2 \left({ \frac{\sqrt{3A}}{2b_0^{9/8}}}\left(\theta -\frac{A}{2b_0^{7/4}} \tau_0\right)\right)
\end{equation}

\noindent where $A$ is a constant amplitude. For $\tau  \geq 0$ this of course is not the  1- soliton solution for the new depth, and we are interested in investigating the changes in the behaviour triggered at $\tau =0$. With rescaling
$$F =-\frac{3}{2} b^{-9/4} E$$ (and appropriate rescaling of $\tau \rightarrow \tau'$) the KdV-type equation \eqref{KdVt1} can be written as a KdV in  a canonical form \begin{equation} F_{\tau'}-6 FF_\theta + F_{\theta \theta \theta}=0 \label{KdV0} \end{equation} with initial condition taken as the one soliton solution for depth $b_0$ at $\tau=\tau'=0:$

$$F(\theta, 0)=-\frac{3}{2} b^{-9/4} E_-(\theta,0)=-\frac{3}{2} b^{-9/4} A \text{sech} ^2 { \frac{\sqrt{3A}}{2b_0^{9/8}}}\theta.$$

\noindent Let us introduce the constant $$K= { \frac{\sqrt{3A}}{2b_0^{9/8}}} \Rightarrow A=\frac{4K^2 b_0^{9/4}}{3}$$ then the initial condition acquires the form

\begin{equation} F(\theta, 0)=-2\left(\frac{b}{b_0} \right)^{-9/4}K^2 \text{sech} ^2 K \theta. \label{ic} \end{equation}
Recall that the initial condition for the one-soliton is $F_{1s}(\theta,0)= -2K^2 \sech^2 (K\theta), $ therefore this initial condition can produce several solitons, waves of radiation, or, in any case can change the shape of the incoming soliton.

For $\tau >0$ the 1-soliton solution is of the form

$$ E_{1s}(\theta,\tau)=A \text{sech} ^2 \left({ \frac{\sqrt{3A}}{2b^{9/8}}}\left(\theta -\frac{A}{2b^{7/4}} \tau\right)\right)$$ where $ b\approx h_0(1-\alpha).$



\noindent Let us denote the constant $$ B=\left( \frac{b}{b_0} \right)^{-9/4}=\left( \frac{1+\alpha}{1-\alpha} \right)^{9/4}. $$

\noindent It is important to know how many solitons ($N$) will be born with the initial condition \eqref{ic}. Clearly, $N$ will be a function of the threshold $\alpha$. (In the case $\alpha=0$ we have obviously $B=1$ and $N=1$.)

For \eqref{KdV0} the associated spectral problem is \cite{Lamb, ZMNP} $$-\psi_{\theta \theta}+ (F(\theta, \tau')-\zeta) \psi=0,$$

\noindent where $\zeta$ is a spectral parameter, which however is time-independent. Therefore it is sufficient to study the spectral problem at $\tau'=0:$ 
$$-\psi_{\theta \theta}+ (F(\theta, 0)-\zeta) \psi=0$$ where $F(\theta,0)= -2BK^2 \sech^2 (K\theta)$ has been given in \eqref{ic}. Change of variables $z=\tanh (K\theta),$  $z\in [-1,1]$ leads to the associated Legendre equation for $\psi$

$$ \frac{d}{dz}\left[ (1-z^2)\frac{d\psi}{dz}\right]+ \left({2}{B}+ \frac{1}{K^2 }\frac{\zeta}{1-z^2} \right) \psi=0. $$

\noindent It has $L^2$ solutions on the interval $[-1,1]$ if

$${2}{B}=N(N+1), \qquad \zeta=-m^2 K^2$$  where $ N$ is an integer, the number of the discrete eigenvalues of the spectral problem and thus the number of the solitons.  $m$ is another integer, labelling the negative discrete eigenvalues, $|m|\leq N$.  The solutions $\psi=P_N^m(z)$ to this equation are called the associated Legendre polynomials \cite{AS}. Therefore there are special depths ({\it eigendepths}) leading to the appearance of $N$ solitons for $\tau>0$:

\begin{equation} B=\frac{N(N+1)}{2}=\left( \frac{b}{b_0} \right)^{-9/4} \Rightarrow  \frac{b}{b_0}= \left(\frac{N(N+1)}{2} \right)^{-4/9}, \label{N} \end{equation} a result that appears in \cite{TZ, Johnson71}. The corresponding values for $\alpha$ are given in Table \ref{Tab1}.

\begin{figure}[h!]
\begin{center}
\fbox{\includegraphics[totalheight=0.23\textheight]{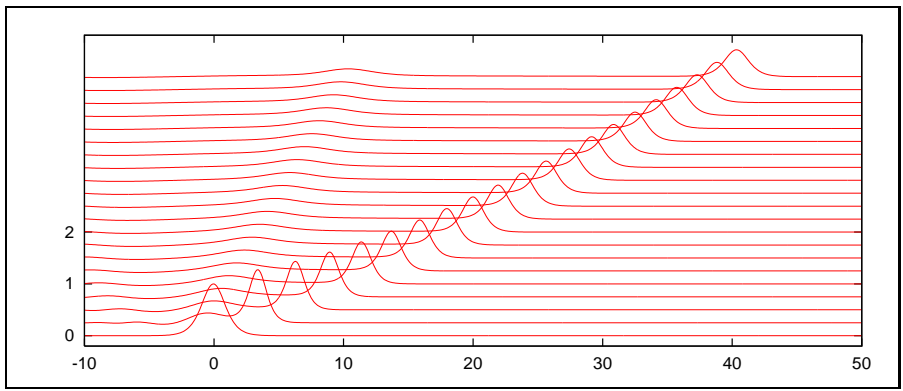}}
\caption{$\alpha=0.3$ with two emerging solitons ($\beta=20, $ $A=1$). A small amplitude (and velocity) soliton is visible behind the bigger one.The horizontal axis is $\theta$ and the snapshots are taken for different $\tau$ in relative units.}
\label{fig:al=03}
\end{center}
\end{figure}

\begin{figure}[h!]
\begin{center}
\fbox{\includegraphics[totalheight=0.23\textheight]{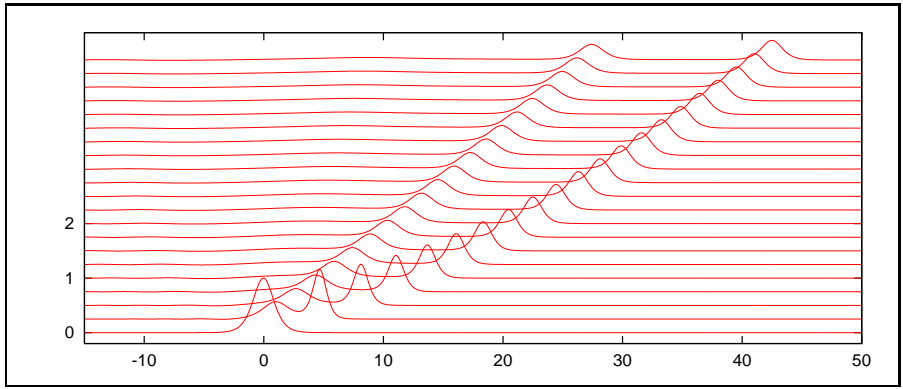}}
\caption{$\alpha=0.5$ with two solitons. The amplitude of the slower soliton is bigger in comparison to the situation in Fig. \ref{fig:al=03}. ($\beta=20, $ $A=1$).}
\label{fig:al=05}
\end{center}
\end{figure}

\begin{figure}[h!]
\begin{center}
\fbox{\includegraphics[totalheight=0.23\textheight]{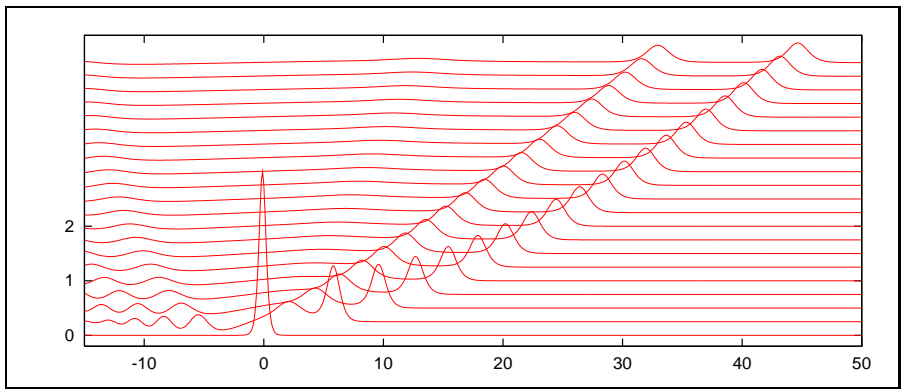}}
\caption{$\alpha=0.5$ with higher initial amplitude $A=3$. The number of solitons in comparison to the situation on Fig. \ref{fig:al=05} is unchanged, however the reflected waves (waves of radiation) are visible on the left, $\beta=20.$}
\label{fig:al=05A3}
\end{center}
\end{figure}

\begin{figure}[h!]
\begin{center}
\fbox{\includegraphics[totalheight=0.23\textheight]{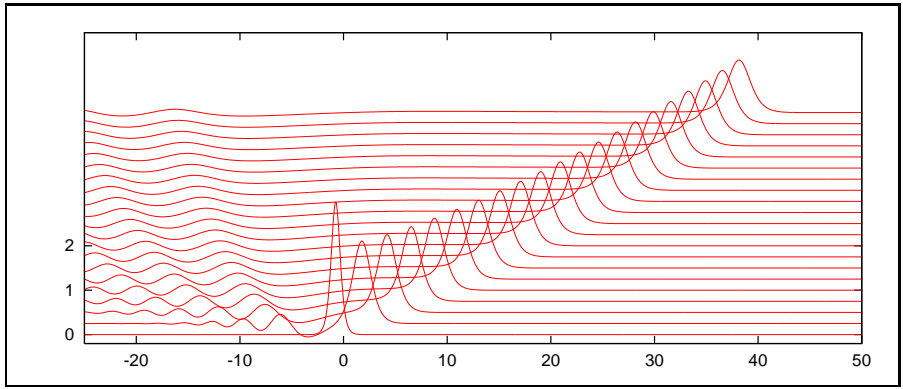}}
\caption{$\alpha=-0.5$ with initial amplitude $A=3$. This corresponds to soliton propagation from shallow water to deep water. There are no new solitons emerging - the energy of the incoming soliton is transformed to energy of the reflected wave only (which is a rapidly decaying wave of radiation), $\beta=20. $ }
\label{fig:al=m05A3}
\end{center}
\end{figure}

We study numerically the evolution of an incoming soliton
\begin{equation}  E_-(\theta,\tau)=A \text{sech} ^2 \left({ \frac{\sqrt{3A}}{2b_0^{9/8}}}\left(\theta -\frac{A}{2b_0^{7/4}} \tau\right)\right) , \qquad \tau<0  \label{incs}
\end{equation}

\noindent which enters a new depth at $\tau=0$ and $\theta=0$ and can transform into a multi-soliton solution for the new depth $b$ (if $b<b_0,$ respectively $\alpha>0$). To this end we consider the fully implicit finite-difference implementation of \eqref{tdkdv} complemented by an inner iteration with respect to the nonlinear term (for more details see, for example \cite{dcds2007}). In Fig. \ref{fig:al=03}, where $\alpha=0.3$, and above the value necessary for the emergence of the two soliton solution (Table \ref{Tab1}) the birth of the second soliton (of a much smaller amplitude) is visible. The difference between $b_0$ and $b$ increases with $\alpha$ and so does the amplitude and the velocity of the second soliton, e.g. when $\alpha=0.5$ for the propagation from Fig. \ref{fig:al=05}. The increase in the amplitude of the incoming soliton only increases the reflected waves which are waves of radiation.   They are highly unstable and decay rapidly with $\tau$,
which can be seen from Fig. \ref{fig:al=05A3}. Note that if the incoming soliton moves from shallow to deep region ($b>b_0$ and $\alpha<0$) then new solitons do not appear, due to \eqref{N}. Then only reflected waves of radiation reduce the energy of the incoming soliton. This is illustrated in Fig. \ref{fig:al=m05A3}.
Note that the dispersive radiation waves of small amplitude move to the left. This is because their phase velocity determined from $ E_{\tau} + \frac{1}{6}\sqrt{b}E_{\theta \theta \theta}=0 $ is $c(k)= - \frac{\sqrt{b}}{6} k^2 <0$ and becomes significant for the short waves where $k$ is not small. This effect is unphysical since the KdV model does not work as a water-wave model for short waves. The soliton velocity is positive as it can be seen e.g. from \eqref{incs}.

\begin{figure}[h!]
\begin{center}
\fbox{\includegraphics[totalheight=0.23\textheight]{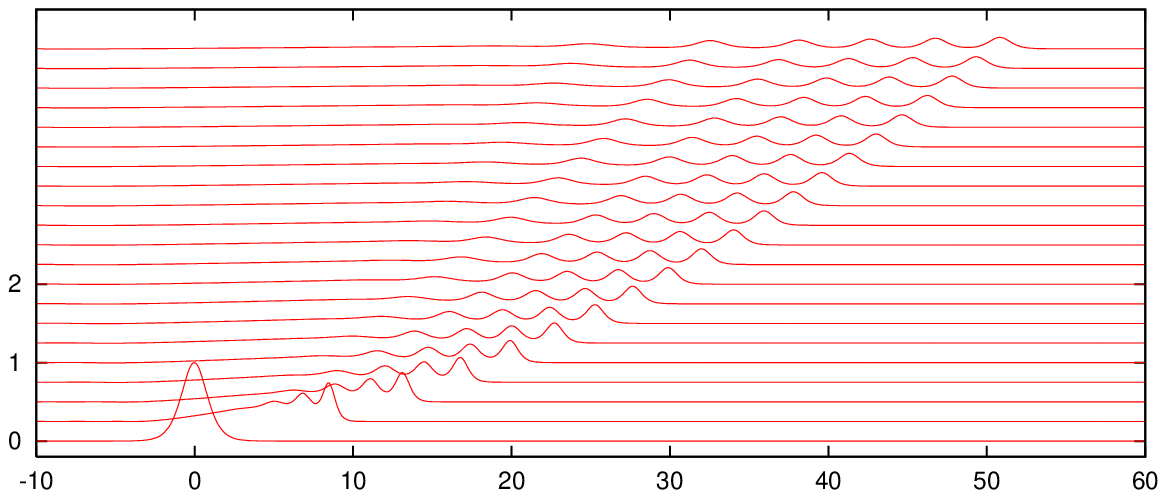}}
\caption{$\alpha=0.8$ with initial amplitude $A=1$. There are several solitons emerging due to the bigger threshold, $\beta=20. $ }
\label{fig:al=08A1}
\end{center}
\end{figure}

 As expected from  \eqref{N} and Table \ref{Tab1} the increase of the threshold $b_0 - b$ (the increase of $\alpha$) increases the number of the emerging solitons, Fig. \ref{fig:al=08A1}. Although qualitatively the numerical results are in an agreement with the theory, the exact values of $N$ from \eqref{N} are not matched. The possible reason is that strictly speaking \eqref{N} is valid for a rapid jump from  $b_0$ to $b$ at $\theta=0, \tau=0$ while our assumption is for slow (and smooth) bottom variations which we model via the $ \tanh$ profile \eqref{b}.

\begin{table}[]
\centering
\caption{The dependence of $N$ on $\alpha.$}
\label{Tab1}
\begin{tabular}{|l|l|l|l|l|l|l|}
\hline
\multicolumn{1}{|c|}{$N$} & 1 & 2 & 3  & 4  & 5  & 6  \\ \hline
$\alpha$                  & 0  &  0.24  & 0.38 & 0.47  & 0.54 & 0.59 \\ \hline
\end{tabular}
\end{table}

\section{Discussion}

The motion of the wave surface is determined by two functions -- the Hamiltonian variables $\eta(x,t)$ and the potential on the surface $\phi(x,t)$. However the fluid motion in the entire domain of the fluid $\Omega$ can only be recovered from the entire boundary of $\Omega$ which includes the bottom. An expression for the continuation of the potential from the surface into the bulk of the fluid is provided in \cite{CraigGuyenneNichollsSulem} in terms of the Dirichlet-Neumann operator. This at least formally determines the dynamic of the fluid in $\Omega$.

 There are of course many other possibilities for the nature of the bottom variation, e.g. random topography studied in \cite{BCDGS,CraigGuyenneSulem}, as well as for the propagation regimes which will be studied in forthcoming publications. The wave dynamics in the presence of shear currents (vorticity) with a variable bottom is another very important and interesting possibility for future research.

\enlargethispage{20pt}


\subsubsection*{Funding} AC and RI acknowledge funding from the Erwin Schr\"odinger International Institute for Mathematics and Physics (ESI), Vienna (Austria) as participants in the Research in Teams Project {\it Hamiltonian approach to modelling geophysical waves and currents with impact on natural hazards}, where a big part of this work has been done. AC is funded by a Fiosraigh fellowship at Dublin Institute of Technology (Ireland). MT acknowledges financial support from the Bulgarian Science Fund under grant DFNI I-02/9.

\subsubsection*{Acknowledgements} The authors are thankful to Prof. Adrian Constantin, Prof. Robin Johnson, Dr Calin I. Martin and Prof. Andr\'e Nachbin for many valuable discussions. The authors are also thankful to two anonymous referees for their very constructive remarks and suggestions.


\end{document}